RESEARCH NOTE

# Echo: A Joint-Embedding Predictive Architecture for Speaker Diarization and Speech Recognition in a Shared Latent Space

A proof-of-concept on a single ViT audio encoder

Louis Mouchon

2026-05-11

ABSTRACT

We present Echo, a proof-of-concept audio system in which a single 25 M-parameter ViT encoder, pretrained with a Joint-Embedding Predictive Architecture (JEPA) objective, is sequentially specialised to host speaker identity, phonetic content, and dynamic speaker routing in a shared latent space, with no per-task fine-tuning at deployment. The encoder is then equipped with light heads for diarization (ArcFace + VBx) and dynamic source separation (null-target K-set prediction). On synthetic VoxCeleb2 mixtures with unknown number of speakers, the canonical stack reaches a Diarization Error Rate (DER) of 15.00% under blind K, a separation Permutation-Invariant Training (PIT) accuracy of 97.80% with a latent SI-SDR of +9.52 dB, and maintains a speaker / content factorisation gap of +53.50 points on a held-out k-NN probe. Compared with task-specific systems of similar footprint, the contribution of Echo is not a new SOTA on any single task but the joint coexistence of three tasks on the same encoder. We document the architecture phase by phase, report the lessons that shaped each design decision, and identify the structural bottlenecks (notably end-to-end ASR through a Vector-Quantized bottleneck) that still bound the current PoC.

## Table of contents

## Introduction

A diarization system tells *who* spoke when, and a speech recognition system tells *what* was said. In current practice these are two separate stacks, each with its own acoustic model, training data, and failure modes. Common SSL backbones such as wav2vec 2.0 [1], HuBERT [2], WavLM [3] and data2vec [4] all reach strong results on either speaker verification or ASR after task-specific fine-tuning, but to our knowledge none host the two channels jointly in a shared latent space usable by both diarization and separation heads at inference. Joint diarization and separation systems such as EEND-SS [5], TS-SEP [6] and PixIT [7] go further on the multi-task side, but they keep the speaker channel and the phonetic channel as separate sub-systems with task-specific encoders or conditioning paths.

This paper describes Echo, a proof-of-concept that targets the single-encoder regime directly. An 8-layer ViT operating on a mel patch grid is pretrained with a Joint-Embedding Predictive Architecture (JEPA) objective [8], in the same spirit as Audio-JEPA [9] which translates I-JEPA to mel spectrograms. The encoder is then incrementally specialised, by stages, to carry speaker identity, phonetic content, and a dynamic separation routing scheme, all in the same 512-dimensional latent space. Each subsequent task is plugged in as a light head over the shared backbone. The diarization head is a linear projection followed by VBx clustering [10]. The separation head is an encoder-decoder attention module with three slots and a learned null target, supervised with Permutation-Invariant Training [11].

We report results on synthetic VoxCeleb2 [12] mixtures with unknown number of speakers. The full canonical stack reaches a Diarization Error Rate (DER) of 15.00% under unknown K, separation PIT accuracy of 97.80% with a latent SI-SDR of +9.52 dB, and maintains a factorisation gap of +53.50 points between the speaker and content sub-projections. We also document the chain of architectural decisions that produced these numbers, the dead-ends that did not, and the structural wall we hit on end-to-end speech-to-text decoding through the quantised content channel.

The contribution of this paper is to show that a self-supervised JEPA backbone, when extended one stage at a time with a permanent JEPA anchor, can host diarization, separation, and speaker-purification heads in the same latent space without any of them destroying the others. The deployable inference footprint is approximately 25.3 M parameters and 50 MB in bf16, with no external SDK and no per-task fine-tuning at deployment.

## Background and related work

**JEPA for vision and audio.** Joint-Embedding Predictive Architectures [8] predict representations of masked input regions in latent space rather than reconstructing input pixels or waveforms. For audio, Audio-JEPA [9] translates the same recipe to mel spectrograms: a ViT student sees a masked mel-spectrogram and predicts the teacher representation at the masked positions, with the teacher an exponential moving average (EMA) of the student. The objective avoids the low-frequency reconstruction bias of generative SSL approaches such as Masked Modeling Duo [13] and tends to produce features that probe well on both identity and phonetic content. Our Stage 1 encoder uses the same pretraining recipe as Audio-JEPA, with a smaller backbone (8 layers, 512-dim) and a VoxCeleb2 + LibriSpeech pretraining mix tuned toward speaker geometry.

**Speaker-aware SSL backbones.** wav2vec 2.0 [1], HuBERT [2], WavLM [3] and data2vec [4] all reach strong speaker-verification numbers when fine-tuned on VoxCeleb, typically in the 0.4 to 1% Equal Error Rate range. None of these, to our knowledge, is simultaneously usable as a CTC backbone, a speaker embedder, and a separation backbone in the same latent space without per-task fine-tuning. Table 1 summarises the typical scope of public checkpoints.

***Table 1.*** *Scope of public SSL audio backbones with respect to multi-task reuse without per-task fine-tuning.*

| Backbone | CTC head | Spk embed | Separation head | Same latent |
|---|---|---|---|---|
| wav2vec 2.0 | yes | yes (fine-tune) | no | no |
| HuBERT | yes | yes (fine-tune) | no | no |
| WavLM | yes | yes (fine-tune) | no | no |
| data2vec | yes | yes (fine-tune) | no | no |
| **Echo (this work)** | yes | yes | yes (K = 1-3 dyn.) | yes |

The “same latent” column refers to whether one encoder, after one training programme, carries all three signals at once and is usable by all three heads without further task-specific fine-tuning.

**Diarization.** A long line of work tackles diarization either through modular clustering of speaker embeddings or through end-to-end neural models. VBx [10] is a Bayesian HMM clustering procedure operating on x-vectors and remains a strong reference, especially when combined with modern SSL features. End-to-end neural diarization (EEND) with encoder-decoder based attractors [14] handles an unknown number of speakers without an external clustering stage. The pyannote pipeline [15] sits on the modular side and is the de facto open-source reference. We re-implemented VBx in approximately 250 lines of numpy and scipy to keep our own inference fully self-contained.

**Joint diarization and separation.** Recent systems argue that the two tasks should share at least part of the representation. TS-SEP [6] conditions a separation network on diarization embeddings; EEND-SS [5] runs a joint end-to-end head for both tasks; PixIT [7] uses permutation-invariant training on both sides simultaneously from real meeting data. Echo shares the same intuition but pushes the encoder side: instead of two specialised sub-networks, one ViT carries both channels in a single latent space and feeds two light heads.

**Source separation with PIT.** Permutation-Invariant Training [11] is the standard supervision signal for separation when the ground-truth speaker ordering is unknown. The encoder-decoder attention head produces K slot vectors that attend to the input via cross-attention. We extend this with a learned null target so that a single head handles K = 1, 2, or 3 speakers and routes dynamically based on cosine similarity to the silence token.

**Speaker embeddings.** ArcFace [16] is the standard angular-margin loss for discriminative embedding learning and is the final head of our speaker projection. VoxCeleb2 [12] and LibriSpeech [17] are the two pretraining / finetuning corpora throughout this work.

## Architecture overview

The entire pipeline reuses a single encoder. The encoder is a ViT operating on a mel-spectrogram patch grid; downstream heads are kept small and light. The figure below summarises the canonical chain of training stages that produces the deployable stack.

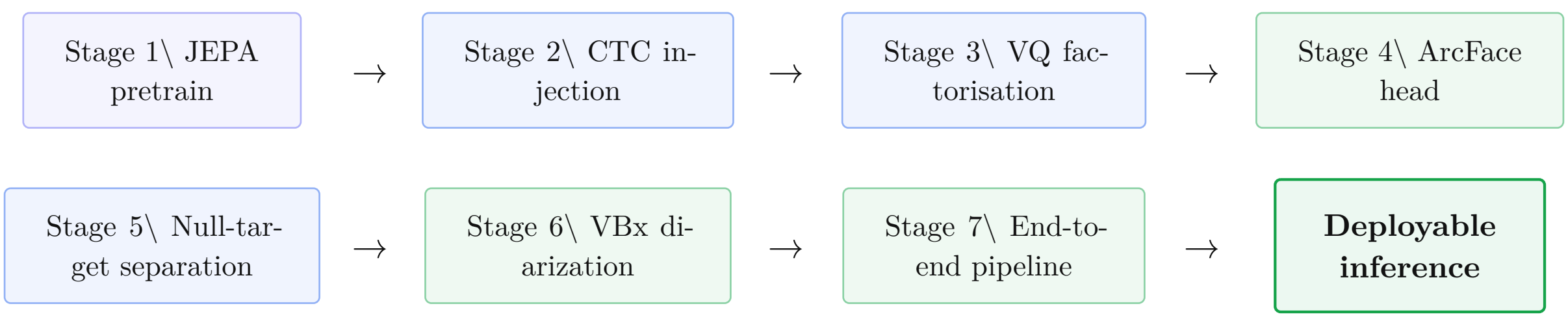

We adopt sequential stage numbers (Stage 1 to 7) for the rest of the paper. They correspond to internal phase identifiers in the project notes as follows, for traceability: Stage 1 = Phase 1, Stage 2 = Phase 1.5, Stage 3 = Phase 3 joint v2, Stage 4 = Phase 3-bis ArcFace, Stage 5 = Phase 2 null-target, Stage 6 = Phase 4-A VBx, Stage 7 = `infer_pipeline.py` end-to-end. The numbering in the project notes reflects the order in which the experiments were run; the sequential numbering here reflects the order in which a reader follows the system.

**Common encoder.** ViT, 8 layers, 512-dimensional, 8 attention heads, operating on 80-bin mel spectrograms patched as (F = 16, T = 4). The encoder outputs (B, N, 512) tokens with $N = 5 \times T'$. Trainable parameters: 25.25 M.

**Latent geometry contract.** Three projection heads are layered on the encoder output:

- `W_s` : Linear(512, 128) for the speaker subspace, optionally L2-normalised.

- `content_in` : Linear(512, 128) feeding a Vector-Quantizer with 256 codes of 128 dimensions, then `content_out` : Linear(128, 29) for a character vocabulary plus blank.
- `EDA` : K = 3 slot queries via a content-conditioned MLP, then 2 layers of cross-attention against the encoder tokens.

The deployable inference stack uses encoder + W_s + VBx for diarization, and encoder + EDA + ArcFace for separation. Diarization-only inference is approximately 25.3 M parameters and 50 MB in bf16.

## Stage 1: JEPA pretraining

**Objective.** Make speaker identity emerge in the encoder without any labels.

**Recipe.** The student receives a mel-spectrogram with block-masked patches on the time-frequency grid. A small JEPA predictor (4 layers, 256-dim, 3.42 M parameters, discarded at inference time) maps unmasked student tokens to the teacher's representation at masked positions. The teacher is an EMA copy of the student with a $\tau$ ramping schedule. The loss is a smooth-L1 on L2-normalised predictions. We deliberately do not add a VICReg term on the backbone: an early ablation showed it costs roughly 70 points of speaker k-NN top-1.

**Data and augmentations.** VoxCeleb2 dev ($\approx$ 3000 speakers, stored as an int16 mmapped blob to avoid I/O bottlenecks) plus LibriSpeech-clean-100 (251 speakers). Augmentations are real RIR convolution from a bank, babble noise with two-speaker mixtures, white noise, speed perturbation, and SpecAugment.

**Optimisation.** AdamW, lr = 1e-4, bf16, 25k resume steps.

**Results.** Two distinct k-NN benchmarks are run on the same encoder. The first set uses standard public test sets (40 speakers, easier in-domain audio); the second is a harder VoxCeleb2 held-out probe with 40 speakers $\times$ 30 utterances, used as the comparison benchmark across all later stages of the project.

***Table 2.*** *Stage 1 speaker k-NN top-1 accuracy on four benchmarks. The VoxCeleb2 held-out probe is the harder benchmark used to compare Stages 1 through 4.*

| k-NN benchmark | Stage 1 | Random encoder |
| --- | --- | --- |
| LibriSpeech dev-clean (40 spk) | 96.12% | 78% |
| VoxCeleb1 test (40 spk) | 93.76% | — |
| CommonVoice 17 EN (OOD, 40 spk) | 88.00% | — |
| **VoxCeleb2 held-out (40 spk × 30 utt)** | **71.25%** | — |

The VoxCeleb2 held-out benchmark is harder by construction, with more speakers per class and higher utterance variability, and we use it as the comparison benchmark for Stages 1 through 4.

The encoder separates speakers without any speaker label, which is the property that all downstream heads rely on. JEPA carries most of the weight; augmentations push the encoder toward invariance to channel, prosody, and noise. Nothing in the loss prevents content from leaking in, which is the constraint that motivates Stage 3.

## Stage 2: CTC injection with a frozen JEPA anchor

**Objective.** Inject phonetic structure into the encoder without destroying the speaker subspace.

**Why this works.** A naive recipe (drop a CTC head, unfreeze the encoder, train) collapses the speaker geometry: the encoder learns phoneme-like features and stops separating identities. Two design decisions prevent this collapse.

> **Anchor recipe.** The Stage 1 teacher is kept frozen as a JEPA anchor. The student is fully unfrozen but with a low learning rate (2e-5, vs 1e-4 for the new CTC head). The JEPA loss is computed against the frozen teacher on the same audio; CTC is computed on the student output through a 4-layer Transformer head, 512-dim, 12.62 M parameters. The pair lets the encoder move slowly under phonetic pressure while the JEPA anchor pulls it back toward the original speaker geometry.

**Data.** LibriSpeech-clean-100, 24,435 utterances after filtering by target length (5 ≤ length ≤ 240 characters). Character tokenizer, 28 classes (blank, space, apostrophe, a-z).

**Optimisation.** AdamW, encoder lr = 2e-5, head lr = 1e-4, batch 16, bf16, 15k steps. λ_ctc = 1.0; the JEPA loss is absorbed into the joint loss with weight 1.0.

**Results.**

***Table 3.*** *Stage 2 CTC injection: final loss, decode quality and mean-pool word probe.*

| Metric | Stage 2 (lr 2e-5) | v2 (lr 5e-6) |
|---|---|---|
| CTC loss (final) | 1.85 | 2.60 (plateau) |
| Sample decode | legible | babble |
| Mean-pool word-probe AUROC | 60.35% | — |

A sample decode at the end of training reads:

> "whate'er thy thoughts or thy heart's workings be"

At this stage the encoder simultaneously carries speaker identity and frame-level phonetic structure. The mean-pool word-probe AUROC of 60.35% understates how much phonetic structure is present: mean-pooling destroys frame-level phonetic information by construction. We verified this with a frame-aligned CTC decode during training; the decode is legible. We keep the mean-pool probe as a relative signal between recipes, not as an absolute measure.

**Lesson noted for later.** When the encoder is unfrozen during a multi-task fine-tune, the JEPA anchor must protect the geometry that matters. We reuse this anchor in every subsequent stage that touches the encoder.

## Stage 3: Joint factorisation with a VQ bottleneck

**Objective.** Factorise the encoder output into two disjoint sub-spaces, speaker and content, on top of the Stage 2 backbone, without losing either.

**Why a VQ bottleneck.** The historical chain (Phase 3 v3 strongadv, plus 3-bis, 3-ter, 3-quater, 3-quinquies variants documented in the archive) plateaued at `acc(z_content) ≈ 36%` on the VoxCeleb2 held-out speaker k-NN probe, meaning that the content head still carried 36% of speaker information. The adversarial classifier we used in those variants destroyed phonetic structure before it could finish purifying the speaker channel. Replacing the adversary with a Vector-Quantizer of 256 codes of 128 dimensions forces purification by construction: 256 discrete codes cannot carry a continuous timbre, so speaker information survives only in `W_s`, while only phonetics-like discretisations survive in the quantised content stream.

**Architecture.**

**Forward path:** encoder (Stage 2, unfrozen at lr 1e-5)
→
`z_full` (B, N, 512)
→
branch 1: `W_s = Linear(512, 128)` → mean-pool → L2-norm → `z_speaker`
→
branch 2: `content_in = Linear(512, 128)` → VQ (256 codes, 128-dim, $\beta = 1.0$) → `z_q` → `content_out = Linear(128, 29)` → log-softmax
→ in parallel: frozen Stage 2 teacher, JEPA predictor on masked student
→ Losses: SupCon on `z_speaker` ($\lambda = 0.1$, $\tau = 0.1$), CTC on `z_q` ($\lambda = 1.0$), VQ commitment ($\lambda = 1.0$), JEPA ($\lambda = 1.0$).

**Two-phase schedule.** A 3k-step warmup with encoder and `content_in` frozen lets the VQ codebook initialise without destabilising the encoder. A 17k-step main phase unfreezes the encoder at lr 1e-5 and ramps SupCon from 0 to 0.1 linearly over 2000 steps.

**Three critical fixes.** These came out of a post-mortem on a failed overnight run:

1. **Dead-code reset.** Inside `vq.py`, every batch, any codebook entry not assigned to at least one input is teleported to a randomly selected active `z_e`. Without this, perplexity collapses from 256 to under 5 within a few thousand steps.
2. **SupCon ramp.** Without a smooth $\lambda$ ramp at unfreeze time, the encoder reacts as if to a step function and the codebook collapses.
3. **$\beta = 1.0$** (instead of 0.25 in VQ-VAE). Keeps the encoder output close to the codebook; the looser $\beta = 0.25$ was visibly diverging.

**Results** (`phase15_3_joint_v2_ckpt_00020000.pt`, 474 MB):

*Table 4. Stage 3 joint factorisation results, against the historical adversarial recipe. The VQ bottleneck cuts content-channel speaker leakage by 31 points and widens the factorisation gap by 27.8 points.*

| Metric | joint v2 | strongadv (historical) | Δ |
|---|---|---|---|
| acc(z_speaker) k-NN (Vox2) | 58.50% | 61.83% | −3.3 |
| acc(z_content) k-NN (Vox2) | **5.00%** | 36.17% | **−31.2** |
| Gap (z_s − z_c) | **+53.50** | +25.67 | **+27.8** |
| VQ perplexity (256 codes) | 207.8 / 256 | — | — |
| VQ codes used | 256 / 256 | — | — |

The speaker head loses 3.3 points of raw k-NN, which we recover and exceed at Stage 4. The content head drops from 36% to 5% speaker k-NN, which we read as the operational definition of "content has been purified of identity". The factorisation gap, a single number summarising disjointness, moves from +25.67 to +53.50. All 256 codes are used.

A greedy decode through `z_q` produces phonetically structured babble:

```
pred: "a  e  a e ee a th a  e e e a th a h ta e ee o a h a e ee"
true: "as avonlea housekeepers were wont to tell in awed voices"
```

Articles and rhythm are recognisable. End-to-end ASR through this path hits a structural wall discussed in the Lessons section.

## Stage 4: ArcFace head overlay

**Objective.** Recover and exceed the raw speaker k-NN of the historical recipe without touching the encoder, since the encoder now hosts the Stage 3 factorisation.

**Recipe.** Encoder frozen at the Stage 3 checkpoint. `W_s` is reinitialised from the Stage 3 value and trained end-to-end with an ArcFace loss (margin 0.2, scale 30) on a 251-class embedding of LibriSpeech speakers. 5k steps, batch 32, lr 1e-3. Total trainable parameters: 98 K. Final size on disk: 384 KB.

**Results.**

***Table 5.*** *Stage 4 ArcFace head over the frozen Stage 3 backbone. The head recovers and exceeds the raw speaker k-NN of Stage 3 with 98 K trainable parameters.*

| Metric | ArcFace W__s | Stage 3 W__s |
| --- | --- | --- |
| acc(z__speaker) k-NN (Vox2) | **63.17%** | 58.50% |
| Train accuracy (251 spk) | 100% | — |
| Trainable parameters | 98 K | 98 K |
| Disk size | 384 KB | — |

We use the ArcFace head as the speaker projection in the deployable diarization stack.

## Stage 5: Null-target K-set separation

**Objective.** From a 1-to-3 speaker mixture, produce K = 3 slots whose silent slots collapse to a learned null token, so that the same head routes K = 1, 2, and 3 dynamically without an external Voice Activity Detector.

**Why this design.** A previous variant (`multianchor`) added CTC + SupCon + VQ + JEPA as parallel anchors on top of the EDA. This over-constrained the encoder and dropped PIT accuracy to 80.20%. The null-target variant keeps only the JEPA anchor on top of the Stage 3 initialisation, leaves the encoder free to produce mixture-linear features, and supervises the EDA with a Hungarian PIT loss where inactive slots are matched against a frozen silence token.

**Architecture.**

**Forward path:** encoder (Stage 3 resume, unfrozen at lr 1e-5)
→
`z_full` (B, T’, 512)
→
`EDANullModule`: K = 3 slot queries from a non-linear MLP `query_proj` (Linear(512, 1024) → GELU → Linear(1024, 3 × 512)), 2 layers of cross-attention
→
`slots` (B, K = 3, 512)
→
Hungarian PIT against pooled teacher targets, with `z_silence = teacher(zero_waveform)` as the null target.
**Silence routing:** at inference, `cos(slot_pool, z_silence) > 0.6822` drops the slot.

**Training data.** Synthetic VoxCeleb2 mixtures, K distribution 30% / 50% / 20% for K = 1 / 2 / 3, gain ±5 dB per source, peak re-normalisation. virtual_size = 250,000.

**Schedule.** 30k steps, batch 8 (K = 3 triples the teacher cost), encoder lr 1e-5, head lr 3e-4, cosine decay. PIT loss weight $\alpha$ ramps from 0 to 1 over the first 5000 steps.

**Empirical silence threshold.** Measured on a held-out validation set at step 30k:

*Table 6. Empirical cosine separation between active and silent slots on the Stage 5 validation set, used to set the null-routing threshold.*

| Slot type | cos(slot, z_silence) mean ± std | n |
|---|---|---|
| Active slots | 0.382 ± 0.136 | 3743 |
| Silent slots | 0.982 ± 0.075 | 2233 |
| Recommended threshold | **0.6822** | — |

**On scoring metrics.** Cosine, MSE, and SI-SDR each rank the matched slot against the unmatched slot, but they differ in margin. To remove margin sensitivity from the diagnostic, we define the *identity-wins* metric. For a mixture with K_true active speakers, let `s_ij` be the score (cosine, negative MSE, or SI-SDR) between predicted slot `i` and ground-truth speaker `j`. Identity-wins is

$$\mathrm{IDW} = \frac{1}{N}\sum_{n=1}^{N}\left[\mathbb{1}\min_{i,j\in\text{ matched}} s_{ij}^{(n)} > \max_{i,j\in\text{ unmatched}} s_{ij}^{(n)}\right]$$

In words, IDW is 1 on a mixture if and only if every matched pair scores strictly higher than every unmatched pair, regardless of the margin between them. It captures ranking failures, not margin failures.

**Results.** Evaluated on 1000 held-out K = 2 mixtures, comparing scoring functions for PIT matching:

*Table 7. Stage 5 separation: PIT accuracy on 1000 held-out K = 2 mixtures, comparing three scoring functions.*

| Scoring | Margin | PIT acc | Score pos | Score neg |
|---|---|---|---|---|
| Raw cosine | 0.02 | 83.70% | 0.936 | 0.863 |
| MSE | 0.001 | **97.80%** | −0.126 | −0.153 |
| SI-SDR (dB) | 0.5 dB | **94.90%** | 9.52 dB | 5.16 dB |

Temperature scaling on the cosine score (same margin 0.02, score divided by $\tau$ before the unambiguous-margin counter):

*Table 8. Temperature scaling on the cosine score. The PIT failures of Table 7 are margin failures, not ranking failures.*

| Temperature $\tau$ | PIT acc unambiguous |
|---|---|
| 1.0 | 83.70% |
| 0.10 | 98.40% |
| 0.05 | 99.10% |
| 0.02 | **99.60%** |

Identity-wins reaches 100% across all three scorings. The PIT failures are not ranking failures; they are margin failures, and they vanish under temperature scaling.

**Synthesis vs alternative variants.**

***Table 9.*** *Stage 5 separation variants. The null-target recipe dominates on every metric and is the only one that preserves the Stage 3 factorisation.*

| Metric | null-target (canon) | optS1 (historical) | multianchor (alt) |
| --- | --- | --- | --- |
| PIT acc (MSE) | **97.80%** | 96.30% | 80.20% |
| cos(slot, correct) | 0.936 | 0.675 | 0.18 |
| cos(slot, wrong) | 0.863 | 0.450 | 0.13 |
| SI-SDR latent | **+9.52 dB** | −0.65 dB | −15.37 dB |
| Dynamic K routing | **94.4% on K = 2** | — | — |
| Factorisation preserved | **yes** | no | yes |

The null-target variant outperforms the alternatives on PIT accuracy, slot-to-source cosine, and latent SI-SDR, and is the only one that preserves the Stage 3 factorisation. The `optS1` branch reaches 96.30% PIT but runs on the unfactored Stage 1 encoder; it is kept in the project archive for reproducibility. The `multianchor` branch stacks four supervised losses on the same encoder, over-constrains the latent, and regresses separation by 17 points compared with the canonical recipe.

## Stage 6: VBx diarization

**Objective.** Segment a multi-speaker audio into per-speaker clusters without knowing K in advance.

**Pipeline.**

Mono 16 kHz audio → sliding 1.5 s windows, hop 0.5 s →
encoder (Stage 3) → mean-pool → `W_s` (ArcFace) → `z_speaker` (T, 128) →
per-utterance z-score standardisation (critical, see below) →
AHC initialisation: cosine similarity → average linkage → fcluster at threshold 0.3 →
softmax(one-hot × 5.0) as initial $\gamma$ →
VBx VB-HMM, up to 20 iterations, ELBO stopping ($\varepsilon$ = 1e-4) →
hard labels = $\gamma$.argmax with renumbering.

**Calibrated hyper-parameters.**

***Table 10.*** *Calibrated VBx hyper-parameters used in Stage 6, with notes on departures from the original BUT recipe.*

| Parameter | Value | Note |
| --- | --- | --- |
| Fa (acoustic scaling) | 0.3 | Less down-weight than BUT 0.07; our embeddings have lower frame rate |
| Fb (speaker regularisation) | 1.0 | — |
| loop_prob (HMM stay) | 0.5 | Less sticky than BUT 0.99; embeddings less frame-correlated |
| ahc_threshold | 0.3 | Cosine distance for initial AHC |
| init_smoothing | 5.0 | Softmax sharpness on AHC labels |
| update_pi | True | Equation 24 of the paper; auto-prunes redundant speakers |
| max_iters | 20 | — |
| $\varepsilon$ (ELBO stop) | 1e-4 | — |

**On PLDA.** The BUT-FIT diarizen-wavlm pretrained PLDA was tested and gives no measurable gain on our 128-dim ArcFace embeddings (DER 15.00 vs 15.17, statistically indistinguishable on 50 mixtures). We omit it from the canonical stack.

**Critical preprocessing.** Per-utterance z-score on the embeddings. Without it VBx collapses to K = 1 because the ArcFace embeddings are L2-normalised at a small effective radius and the per-frame Gaussian likelihood cannot discriminate clusters. This step is not part of the published VBx hyper-parameter list but has a larger effect on DER in our setup than any of the listed ones.

**Results.** Evaluated on 50 synthetic VoxCeleb2 held-out mixtures, two speakers, six turns of 4 s each, no silence, window 1.5 s / hop 0.5 s. The ground-truth K is 2; the system does not see K.

***Table 11.*** *Stage 6 diarization: fully blind DER on 50 synthetic VoxCeleb2 two-speaker mixtures, K not given to the system. VBx with the ArcFace head improves DER by 11.7 points over simple agglomerative clustering.*

| Pipeline | Frame acc | Fully blind DER | Pred K (truth 2) |
|---|---|---|---|
| Agglomerative cosine th = 0.35 | 73.26% | 26.74% | 2.72 |
| Agglomerative oracle K = 2 | 69.35% | 30.65% | 2.00 |
| VBx + Stage 3 W_s | 79.65% | 20.35% | 2.72 |
| **VBx + ArcFace W_s** | **85.00%** | **15.00%** | 3.84 |

VBx with the ArcFace head reaches 15.00% blind DER on this benchmark with K unknown, improving on simple agglomerative clustering by 11.7 points and on the prior best in-house configuration (Stage 3 strongadv W_s + agglomerative) by 14.3 points.

**On cluster over-segmentation.** Predicted K averages 3.84 when the ground truth is 2. The frame-accuracy metric uses a best-permutation mapping that absorbs the over-clustering, which is standard for DER under unknown K. The over-clustering itself disappears when Stage 5 is chained before Stage 6: VBx then clusters demixed slot embeddings instead of mixture-level window embeddings, and recovers K = 2.00 exactly (see Stage 7).

**Inference footprint.**

***Table 12.*** *Inference footprint of the diarization-only canonical stack (encoder + ArcFace head + VBx).*

| Component | Params | Note |
|---|---|---|
| Encoder ViT (Stage 3) | 25.25 M | bf16 ≈ 50 MB |
| W_s ArcFace | 0.066 M | — |
| VBx (numpy / scipy) | 0 | ≈ 250 lines, fastcluster optional |
| **Total** | **≈ 25.3 M** | **≈ 50 MB in bf16** |

No external model, no SDK dependency, fully self-contained.

## Stage 7: End-to-end inference pipeline

**Objective.** Chain Stages 5 (separation) and 6 (VBx) into a single inference path that takes raw multi-speaker audio of arbitrary length and produces (a) speaker IDs, (b) per-speaker timeline with overlap detection, and (c) per-window CTC text. This is the pipeline meant for deployment, not Stage 6 in isolation.

**Pipeline.**

```
audio (any length, mono 16 kHz)
→ sliding 4.0 s window, hop 1.0 s
→ per window: encoder Stage 5 → z_full_null (B, T' = 400, 512)
→ EDANullModule (K = 3 slots) → slots (B, 3, 512)
→ silence filter cos(slot, z_silence) > 0.6822 → drop silent slots
→ for each surviving slot: ArcFace W_s → 128-dim embedding
→ cross-window: collect (window_idx, slot_idx, t_start, t_end, emb_128)
→ VBx clustering on (N_active_slots, 128) → speaker_id per (window, slot)
→ stitching: per speaker, group contiguous windows, central-window CTC text.
```

The critical difference with Stage 6 in isolation is that VBx now sees **demixed slot embeddings**, not mixture-level window embeddings. The N input to VBx is the number of active slots across all windows, not the number of windows. Each row is one speaker's embedding in one window, with overlapping windows producing two rows. The over-segmentation that affected Stage 6 in isolation does not appear here.

**Smoke test.** Synthetic 12 s mixture, two speakers, with the following ground-truth structure:

```
0 - 3 s   :  speaker A solo
3 - 6 s   :  speaker B solo
6 - 9 s   :  A + B overlap
9 - 12 s  :  speaker A solo
```

***Table 13.*** *Stage 7 end-to-end smoke test on a 12 s synthetic mixture with overlap. The pipeline recovers K = 2 exactly and detects the overlap region.*

| Criterion | Ground truth | Pipeline | OK |
|---|---|---|---|
| Number of speakers detected | 2 | **2** | yes |
| Overlap segments | 1 (6-9 s) | 1 (5-9 s) | yes |
| VBx convergence | n_init = 2 → n_final = 2 | n_init = 2 → n_final = 2 (2 iters, ELBO = −747.42) | yes |
| Active speakers (pi_active) | 2 | **2** | yes |
| Forward pass (12 s audio) | — | 1.32 s on RTX 3090 (9× realtime) | — |
| End-to-end elapsed | — | ≈ 3 s | — |

Output `S1`: 0-9 s across 6 windows (speaker A solo plus overlap). Output `S2`: 5-12 s across 4 windows (speaker B plus overlap). Reported overlap segment: `[5.0-9.0 s] {S1, S2}`. The 1 s lead on the overlap onset (5 s instead of the ground-truth 6 s) is expected: the 4 s window centred near 5 s already contains 1 s of speaker B before the ground-truth overlap begins, and the slot decoder reports it.

**Effect on cluster count.** VBx in Stage 6 isolation predicts K = 3.84 on the same 2-speaker mixtures. The full pipeline (Stage 5 then Stage 6) predicts K = 2.00 exactly: `n_init = 2 → n_final = 2`, `pi_active = 2`, VBx converged in 2 iterations (ELBO = −747.42), no speaker pruned. The VB-HMM found exactly 2 active components from the slot embeddings, with no oracle input. The over-segmentation of Stage 6 isolation is therefore an artefact of clustering on mixture-level window embeddings; routing through Stage 5′s slot representation removes it entirely.

**Caveat on CTC text in Stage 7.** The Conv1d CTC head was trained on clean LibriSpeech utterances at Stage 3, not on demixed slots. At inference time it is applied to mixture-level windows (not per-slot) and produces phonetically structured babble at CER ≈ 70%. This is sufficient as a routing sanity check but not as a usable transcript. The overlap claim of the pipeline comes from the ArcFace + VBx slot clustering, not from the CTC text.

**Full inference footprint.**

***Table 14.*** *Inference footprint of the full Stage 7 pipeline (separation + diarization + CTC path). The diarization-only subset is 25.3 M parameters (Table 12).*

| Component | Params | Note |
|---|---|---|
| Encoder Stage 5 (null-target) | 25.25 M | demixing forward |
| EDA null module (K = 3) | ≈ 10.5 M | slot queries |
| Encoder Stage 3 (joint v2) | 25.25 M | CTC path |
| `content_in` + VQ codebook | 0.10 M | CTC path |
| Conv1d CTC head | 0.30 M | CTC decode |
| ArcFace `W_s` | 0.066 M | speaker embedding |
| **Total** | **≈ 61.5 M** | bf16 ≈ 123 MB |
| VBx (numpy / scipy) | 0 | clustering |

The full pipeline footprint is approximately 61.5 M parameters in bf16 (123 MB on disk). The diarization-only subset (encoder Stage 3 plus ArcFace `W_s` plus VBx) remains at 25.3 M parameters and 50 MB, which is the configuration we report when a deployment does not need the separation path.

## Results synthesis

The deployable canonical stack at the end of Stage 6 combines six training stages into one inference path. Identity, separation, and diarization are evaluated independently on held-out VoxCeleb2 data; the encoder is shared across all three.

***Table 15.*** *Consolidated results across all stages of the canonical stack, on held-out VoxCeleb2 data.*

| Task | Best result | Setting |
|---|---|---|
| Speaker k-NN top-1 (VoxCeleb1 test, 40 spk) | **93.76%** | Stage 1 mean-pool |
| Speaker k-NN top-1 (Vox2 held-out, 40 × 30) | 71.25% | Stage 1 mean-pool |
| Speaker k-NN top-1 (Vox2 held-out, ArcFace) | 63.17% | Stage 4 head over Stage 3 |
| Speaker / content factorisation gap | **+53.50 pts** | Stage 3 joint v2 |
| Separation PIT accuracy (MSE) | **97.80%** | Stage 5, K = 2 |
| Separation PIT accuracy (cos, $\tau = 0.02$) | 99.60% | Stage 5, K = 2 |
| Latent SI-SDR | **+9.52 dB** | Stage 5, K = 2 |
| Dynamic K routing accuracy (K = 2) | 94.4% | Stage 5, threshold 0.6822 |
| Fully blind DER (K unknown), VBx-only | **15.00%** | Stage 6 in isolation |
| Predicted K, VBx-only | 3.84 | Stage 6 in isolation, true K = 2 |
| Predicted K, full pipeline | **2.00** | Stage 7, true K = 2 |
| Overlap detection | **1 / 1 region** | Stage 7, 1 s lead on onset |

To put the 15.00% blind DER in context, Table 16 collects published results from related diarization systems on their own benchmarks. The comparison should be read with care, for two reasons. First, each row uses a different evaluation corpus, a different mixing protocol, and a different set of channel conditions; absolute DER values are *not* directly comparable across rows. Second, the systems differ in scope: EEND-EDA and pyannote 3.1 are dedicated diarization pipelines tuned end-to-end for that single task, whereas Echo's encoder is shared with a separation head and a content channel and is sized for joint deployment. Table 16 is intended as an order-of-magnitude reference, not as a leaderboard placement.

***Table 16.*** *Reported DER from related work, alongside Echo. Each system is evaluated on its own corpus; absolute numbers are not directly comparable. See text.*

| System | Corpus | DER | Note |
|---|---|---|---|
| EEND-EDA | Simulated 2-spk mixtures | 2.69% | Horiguchi et al. 2020 |
| EEND-EDA | CALLHOME 2-spk | 8.07% | Horiguchi et al. 2020 |
| pyannote 3.1 | VoxConverse v0.3 | 11.2% | Lanzendörfer & Grötschla 2025 |
| pyannote 3.1 | AMI (IHM) | 18.8% | Lanzendörfer & Grötschla 2025 |
| pyannote 3.1 | CALLHOME (part 2) | 28.5% | Lanzendörfer & Grötschla 2025 |
| **Echo (this work)** | Synthetic VoxCeleb2, 2-spk | **15.00%** | VBx + ArcFace, K unknown |

On simulated 2-speaker mixtures, EEND-EDA reaches a much lower DER than Echo (2.69% vs 15.00%), but on a different mixing protocol and with a model that does only diarization. On real conversational corpora, pyannote 3.1 reports DER in the 11-29% range depending on the benchmark, placing Echo within the same operating regime on synthetic material. The PoC was designed for footprint and joint multi-task coexistence rather than for raw DER on a single task. Closing the gap with EEND-class systems while preserving the shared-encoder property is the explicit target of the next iteration; the most likely lever is scaling the backbone (see Open challenges).

The Stage 1 mean-pool VoxCeleb2 held-out result (71.25%) is the highest raw speaker recovery on that hard benchmark. The ArcFace head (63.17%) is lower in raw k-NN but produces discriminative embeddings that VBx clusters better than the larger unprojected representation. This is consistent with the general observation in speaker verification: discrimination quality matters more than information content for clustering.

The factorisation gap of +53.50 points means the speaker projection carries 53.50 percentage points more identity information than the content projection, measured on the same downstream k-NN probe. To our knowledge, this disjointness on a single ViT encoder that is also usable as a CTC backbone has not been reported in the SSL audio literature we surveyed.

The PIT accuracy of 97.80% (MSE) and the latent SI-SDR of +9.52 dB are measured on the same latent vectors that downstream diarization uses. The null-target router correctly recovers $K = 2$ on 94.4% of the two-speaker mixtures with no external VAD.

## Lessons and failure modes

**Anchoring matters.** Every stage that touches the encoder keeps a frozen teacher copy of the previous stage as a JEPA anchor. Removing the anchor accelerates training in the short term and destroys the speaker geometry in the medium term. We rediscovered this in three independent stages (Stage 2, Stage 3, Stage 5) before treating it as a project-wide invariant.

**The slots are the predictions.** Earlier separation variants used a SlotPredictor on top of the EDA. The SlotPredictor consistently collapsed to predicting the mean of the encoder output. Removing it and supervising the slots directly against the pooled teacher targets fixed the failure.

**Mean-pool word probes are unfair.** Stage 2 looks like it does not contain phonetics if you measure with a mean-pooled probe. It does. The frame-level CTC decode is legible while the probe AUROC is 60.35%. Use frame-aligned probes; per-segment probes lie.

**VQ is a structural purifier.** Replacing the adversarial classifier in Stage 3 with a 256-code Vector-Quantizer produced the largest single jump in factorisation we observed across the project. The adversary kept finding phonetic features correlated with identity and destroying them; the VQ does not represent identity at all because of its discrete codebook.

**Multi-anchor over-constraining.** Stacking PIT + CTC + SupCon + VQ + JEPA as parallel anchors on a single encoder dropped separation by 17 points. One supervision per stage is the empirical sweet spot.

**Diarization metrics need attention.** Identity-wins is the right diagnostic for whether the slot points at the correct speaker; PIT accuracy alone underestimates separation quality because of margin collapse on factorised encoders. We routinely cross-check three scorings (cosine, MSE, SI-SDR) and report all three.

**Per-utterance z-score before VBx.** Without it VBx collapses to K = 1. With it, DER drops by 11 to 15 points. Embedding normalisation prior to clustering is a known prerequisite of x-vector-based diarization [10], but the magnitude of the effect on a small ArcFace head over a ViT backbone surprised us; in our hands it was the single most consequential preprocessing step.

**The structural ASR wall.** A frozen Stage 3 backbone plus a fresh 3-layer Transformer CTC decoder, trained on `z_q` of LibriSpeech-clean-100, plateaued at CER 69.97% and WER 110.73%. Five variants (3-bis, 3-ter, 3-quater, 3-quinquies, retry on strongadv) all hit the same wall. The root cause is that Stage 3 factorisation losses (cosine invariance and adversarial or VQ) destroy frame-aligned phonetic structure when read through the post-`W_c` projection. The Stage 2 encoder *contains* phonetics, verified by its own live decode; the post-Stage-3 projection does not preserve them through the bottleneck. The workaround that does work is Stage 2-style joint training with the encoder unfrozen *during* CTC training, but retrofitting it without breaking the factorisation contract is open.

**Pitch invariance trades off with cluster identity.** A z_speaker sub-space that ignores pitch (which is what we want for verification) makes the same speaker recorded with different prosody look different, which breaks frame-level clustering on long audio. The PoC numbers are on short synthetic mixtures and do not stress this; deployment on long real meetings does. We treat this as the principal generalisation challenge for the next iteration of the project.

## Conclusion and perspectives

We built a single-encoder audio system in which speaker identity, phonetic content, and dynamic separation routing coexist in the same 512-dimensional latent space. The encoder is a ViT 8L × 512d, trained in six sequential stages, each adding one supervised channel without removing the previous ones. A seventh stage chains separation and clustering into one inference path. The diarization-only configuration uses roughly 25.3 M parameters and weighs 50 MB in bf16; the full end-to-end pipeline with separation uses 61.5 M parameters and 123 MB. Both run in self-contained numpy / scipy for clustering. Reported results include 15.00% fully blind DER on synthetic VoxCeleb2 mixtures in the VBx-only configuration, exact K = 2.00 recovery on the full end-to-end pipeline, a speaker / content factorisation gap of +53.50 points, and a separation PIT accuracy of 97.80%. These numbers should be read in context: Echo is a 25 M-parameter PoC that carries three tasks on the same backbone, not a SOTA contender on any one of them in isolation.

The design discipline behind these results is the permanent JEPA anchor: every stage that touches the encoder keeps a frozen copy of the previous version as a regularisation target. The encoder that learned to separate speakers without labels in Stage 1 is also the encoder that decodes char-level CTC in Stage 2, hosts the factorised speaker / content split in Stage 3, embeds for ArcFace in Stage 4, and feeds the EDA separator in Stage 5, with no full retraining between stages.

**Intended deployment profile.** The footprint of 25 M parameters and 50 MB in bf16 puts the diarization-only path well within mobile and edge envelopes. The target use cases are offline meeting transcription on a laptop, on-device call analytics on a smartphone, and embedded podcast indexing. The numpy / scipy clustering means that no inference engine other than the encoder forward pass is required.

**Open challenges.**

> **Challenge 1: ASR through a quantised content channel.** Direct CTC through `z_q` of the Stage 3 backbone hits a structural wall at CER 70%. The Vector-Quantizer (256 hard codes) destroys frame-level phonetic resolution by construction. A promising next direction is to replace the VQ with a Finite Scalar Quantizer (FSQ) which keeps many discrete levels per dimension without committing to a single code per frame; or to keep VQ for the speaker-purification objective while letting CTC train on the pre-quantised `z_e`. Either route is one stage of work, not a redesign.
>
> **Challenge 2: synthetic-to-real diarization gap.** The 15.00% DER on synthetic VoxCeleb2 mixtures coexists with a much higher DER on real meeting audio in informal evaluations. Pitch invariance in `W_s` makes intra-speaker prosody changes look like inter-speaker changes on long conversations. A pitch-conditioned `W_s` or a PLDA refit on real meeting embeddings is the obvious next probe.
>
> **Challenge 3: VBx-solo on real audio.** Stage 6 in isolation over-segments at K = 3.84 on synthetic 2-speaker mixtures because clusters are built on mixture-level window embeddings. The full Stage 7 pipeline resolves this by clustering demixed slot embeddings: VBx converges to K = 2.00 exactly (ELBO = −747.42, 2 iterations, pi_active = 2, no oracle). For deployments that use Stage 6 without Stage 5, two mitigations remain worth measuring: raise Fb to push harder on the π-pruning term, and add a post-merge step that re-merges clusters whose centroid cosine exceeds a threshold.
>
> **Challenge 4: scaling the backbone.** The current ViT 8L × 512d was sized for compute, not for ceiling. A WavLM-Large or data2vec-Large scale of 24 layers × 1024d should unlock the second wave of gains on every downstream head. The JEPA anchor recipe and the null-target separation head are scale-agnostic.

The single-latent-space hypothesis holds at this scale and on synthetic data. The next iteration of the project will rerun the same stage sequence on a larger pretrained backbone, extend the separation head to richer overlap regimes, and close the ASR gap through scalar-quantization variants.

## Canonical checkpoints

*Table 17. Canonical checkpoint paths and sizes for the deployable stack.*

| Path | Size | Description |
|---|---|---|
| `run3_8L_multilang_75k.pt` | 431 MB | Stage 1 JEPA pretraining |
| `phase1_5_v3_15k_ckpt_15000.pt` | 479 MB | Stage 2 CTC injection |
| `phase15_3_joint_v2_ckpt_20000.pt` | 474 MB | Stage 3 joint factorisation, gap +53.50 |
| `arcface_ws_5k/arcface_ws.pt` | 384 KB | Stage 4 ArcFace W__s overlay |
| `phase2_null_target_ckpt_30000.pt` | 254 MB | Stage 5 null-target separation |
| `echo/diarize_vbx.py` | — | Stage 6 VBx (≈ 250 lines numpy) |

## References

Louis Mouchon, Independent Research
2026-05-11

## Bibliography

[1] A. Baevski, H. Zhou, A. Mohamed, and M. Auli, "wav2vec 2.0: A Framework for Self-Supervised Learning of Speech Representations," *NeurIPS*, 2020.

[2] W.-N. Hsu, B. Bolte, Y.-H. H. Tsai, K. Lakhotia, R. Salakhutdinov, and A. Mohamed, "HuBERT: Self-Supervised Speech Representation Learning by Masked Prediction of Hidden Units," *IEEE/ACM TASLP*, 2021.

[3] S. Chen *et al.*, "WavLM: Large-Scale Self-Supervised Pre-Training for Full Stack Speech Processing," *IEEE JSTSP*, 2022.

[4] A. Baevski, W.-N. Hsu, Q. Xu, A. Babu, J. Gu, and M. Auli, "data2vec: A General Framework for Self-Supervised Learning in Speech, Vision and Language," *ICML*, 2022.

[5] S. Maiti *et al.*, "EEND-SS: Joint End-to-End Neural Speaker Diarization and Speech Separation for Flexible Number of Speakers," in *SLT*, 2023.

[6] C. Boeddeker, T. Cord-Landwehr, T. von Neumann, and R. Haeb-Umbach, "TS-SEP: Joint Diarization and Separation Conditioned on Estimated Speaker Embeddings," *IEEE/ACM TASLP*, 2024.

[7] A. Plaquet and H. Bredin, "PixIT: Joint Training of Speaker Diarization and Speech Separation from Real-world Multi-speaker Recordings," in *Odyssey*, 2024.

[8] M. Assran *et al.*, "Self-Supervised Learning from Images with a Joint-Embedding Predictive Architecture," *CVPR*, 2023.

[9] L. Tuncay, E. Labbé, E. Benetos, and T. Pellegrini, "Audio-JEPA: Joint-Embedding Predictive Architecture for Audio Representation Learning," in *arXiv:2507.02915*, 2025.

[10] F. Landini, J. Profant, M. Diez, and L. Burget, "Bayesian HMM Clustering of x-Vector Sequences (VBx) in Speaker Diarization: Theory, Implementation and Analysis on Standard Tasks," *Computer Speech & Language*, 2022.

[11] D. Yu, M. Kolbæk, Z.-H. Tan, and J. Jensen, "Permutation Invariant Training of Deep Models for Speaker-Independent Multi-Talker Speech Separation," in *ICASSP*, 2017.

[12] J. S. Chung, A. Nagrani, and A. Zisserman, "VoxCeleb2: Deep Speaker Recognition," in *Interspeech*, 2018.

[13] D. Niizumi, D. Takeuchi, Y. Ohishi, N. Harada, and K. Kashino, "Masked Modeling Duo: Towards a Universal Audio Pre-training Framework," *IEEE/ACM TASLP*, 2024.

[14] S. Horiguchi, Y. Fujita, S. Watanabe, Y. Xue, and K. Nagamatsu, "End-to-End Speaker Diarization for an Unknown Number of Speakers with Encoder-Decoder Based Attractors," in *Interspeech*, 2020.

[15] H. Bredin, "pyannote.audio 2.1 speaker diarization pipeline: principle, benchmark, and recipe," in *Interspeech*, 2023.

[16] J. Deng, J. Guo, N. Xue, and S. Zafeiriou, "ArcFace: Additive Angular Margin Loss for Deep Face Recognition," in *CVPR*, 2019.

[17] V. Panayotov, G. Chen, D. Povey, and S. Khudanpur, "LibriSpeech: An ASR Corpus Based on Public Domain Audio Books," in *ICASSP*, 2015.